%% file: DAFx25_arxiv_v2.tex
\def\papertitle{Unsupervised Estimation of Nonlinear Audio Effects:\\ Comparing Diffusion-Based and Adversarial approaches}

\def\paperauthorA{Eloi Moliner}
\def\paperauthorB{Michal Švento}
\def\paperauthorC{Alec Wright}
\def\paperauthorD{Lauri Juvela}
\def\paperauthorE{Pavel Rajmic}
\def\paperauthorF{Vesa Välimäki}

\documentclass[twoside,a4paper]{article}
\usepackage{etoolbox}
\usepackage{dafx_25}
\usepackage{amsmath,amssymb,amsfonts,amsthm}
\usepackage{euscript}
\usepackage[T1]{fontenc}
\usepackage[utf8]{inputenc}
\usepackage{ifpdf}
\usepackage[english]{babel}
\usepackage{lipsum}
\usepackage{caption}
\usepackage{color}
\usepackage{booktabs}
\usepackage[nolist]{acronym}

\usepackage[inkscape,inkscapelatex=false]{svg}

\usepackage{float}
\usepackage{tikz}
\usetikzlibrary{positioning, backgrounds, fit, calc, patterns, arrows.meta, shapes.geometric, shapes.arrows}
\usetikzlibrary{arrows.meta}
\usepackage{subcaption}
\usepackage[skins]{tcolorbox}
\usepackage{adjustbox}

\usepackage{algorithm}
\usepackage{algpseudocodex}

\input glyphtounicode
\ninept

\newcounter{numauth}
\newcounter{listcnt}
\newcommand\authcnt[1]{\ifdefined#1 \stepcounter{numauth} \fi}

\newcommand\addauth[1]{
\ifdefined#1 
\stepcounter{listcnt}
\ifnum \value{listcnt}<\value{numauth}
\appto\authorslist{, #1}
\else
\appto\authorslist{~and~#1}
\fi
\fi}
\authcnt{\paperauthorB}
\authcnt{\paperauthorC}
\authcnt{\paperauthorD}
\authcnt{\paperauthorE}
\authcnt{\paperauthorF}
\authcnt{\paperauthorG}
\authcnt{\paperauthorH}
\authcnt{\paperauthorI}
\authcnt{\paperauthorJ}
\def\authorslist{\paperauthorA}
\addauth{\paperauthorB}
\addauth{\paperauthorC}
\addauth{\paperauthorD}
\addauth{\paperauthorE}
\addauth{\paperauthorF}
\addauth{\paperauthorG}
\addauth{\paperauthorH}
\addauth{\paperauthorI}
\addauth{\paperauthorJ}

\usepackage{times}

\newif\ifpdf
\ifx\pdfoutput\relax
\else
   \ifcase\pdfoutput
      \pdffalse
   \else
      \pdftrue
   \fi
\fi

\ifpdf %
  \usepackage[pdftex,
    pdftitle={\papertitle},
    pdfauthor={\authorslist},
    pdfsubject={Proceedings of the 28th International Conference on Digital Audio Effects (DAFx25)},
    colorlinks=false, %
    bookmarksnumbered, %
    pdfstartview=XYZ %
  ]{hyperref}
\else %
  \usepackage[dvips]{epsfig,graphicx}
  \usepackage[dvips,
    pdftitle={\papertitle},
    pdfauthor={\authorslist},
    pdfsubject={Proceedings of the 28th International Conference on Digital Audio Effects (DAFx25)},
    colorlinks=false, %
    bookmarksnumbered, %
    pdfstartview=XYZ %
  ]{hyperref}
  
\fi
\usepackage[hypcap=true]{caption}
\title{\papertitle}

\affiliation{
\paperauthorA $^{\ast 1}$,
\paperauthorB $^{\ast 2}$,
\paperauthorC $^{3}$,
\paperauthorD $^{1}$,
\paperauthorE $^{2}$ and
\paperauthorF $^{1}$
\thanks{\vspace{-3mm}}\hspace{-2pt}
}
{
$^1$\href{https://www.aalto.fi/en/aalto-acoustics-lab}{Acoustics Lab}, Department of Information and Communications Engineering, Aalto University, Espoo, Finland \\
$^2$Department of Telecommunications, FEEC, Brno University of Technology, Brno, Czech Republic \\
$^3$Acoustics and Audio Group, University of Edinburgh, Edinburgh, UK \\
$^{\ast}$Equal contribution \hspace{0.5cm} 
{\tt \href{mailto:eloi.moliner@aalto.fi}{eloi.moliner@aalto.fi}
\hspace{0.5cm}
\href{mailto:michal.svento@vut.cz}{michal.svento@vut.cz}
}
}

\begin{document}
\ifpdf %
  \DeclareGraphicsExtensions{.png,.jpg,.pdf}
\else  %
  \DeclareGraphicsExtensions{.eps}
\fi

\begin{acronym}[]
    \acro{wh}[W-H]{Wiener-Hammerstein}
    \acro{lti}[LTI]{Linear-Time Invariant}
    \acro{fad}[FAD]{Fr{\'e}chet Audio Distance}
    \acro{em}[EM]{Expectation-Maximization}
    \acro{cnn}[CNN]{Convolutional Neural Network}
    \acrodefplural{cnn}[CNNs]{Convolutional Neural Networks}
    \acro{rnn}[RNN]{Recurrent Neural Network}
    \acrodefplural{rnn}[RNNs]{Recurrent Neural Networks} 
    \acro{gan}[GAN]{Generative Adversarial Network}
    \acrodefplural{gan}[GANs]{Generative Adversarial Networks} 
    \acro{dps}[DPS]{Diffusion Posterior Sampling}
    \acro{mss}[MSS]{MultiScale Spectrogram}
    \acro{stft}[STFT]{Short-time Fourier Transform}
    \acro{gcn}[GCN]{Gated Convolution Network}
    \acro{ccr}[CCR]{Cubic Catmull-Rom}
    \acrodefplural{gcn}[GCNs]{Gated Convolutional Networks} 
\end{acronym}

\maketitle

\begin{abstract}
Accurately estimating nonlinear audio effects without access to paired input-output signals remains a challenging problem.
This work studies unsupervised probabilistic approaches for solving this task.
 We introduce a method, novel for this application, based on diffusion generative models for blind system identification, enabling the estimation of unknown nonlinear effects using black- and gray-box models.
 This study compares this method with a previously proposed adversarial approach, analyzing the performance of both methods under different parameterizations of the effect operator and varying lengths of available effected recordings.
Through experiments on guitar distortion effects, we show that the diffusion-based approach provides more stable results and is less sensitive to data availability, while the adversarial approach is superior at estimating more pronounced distortion effects.
Our findings contribute to the robust unsupervised blind estimation of audio effects, demonstrating the potential of diffusion models for system identification in music technology.

\end{abstract}

\section{Introduction}
\label{sec:intro}

Audio effects play a pivotal role in shaping the timbral characteristics of music and audio signals.
Systems such as guitar amplifiers, dynamic range compressors, and fuzz pedals introduce complex nonlinear transformations that define their sonic signature. Emulating these transformations with software has important applications in music production.
When paired input-output recordings are available, supervised learning methods can effectively approximate these transformations \cite{wright2020real, martinez2020deep}.
However, in many practical scenarios, obtaining paired data is challenging or infeasible. This occurs when attempting to reproduce the effects processing used in a recording or when the target device is unavailable.
Such a~situation necessitates blind estimation, where the effect must be inferred without access to the dry input signal, making it a highly underdetermined problem due to the unknown source.

Existing approaches are primarily data-driven, with most relying on supervised learning and pre-training deep neural networks on diverse effect populations, assuming they will generalize to unseen systems.
Since paired data can often be generated on-the-fly by randomizing effect parameters, some of these methods are classified as self-supervised. For example, prior works trained models to infer effect parameters from processed audio \cite{steinmetz2022style, peladeau2024blind} or predict signal chains \cite{lee2023blind, take2024audio, rice2023general}.
Other approaches use contrastive learning to determine representations that extract effect information \cite{chen2024towards, wright2024open, steinmetz2024stito, koo2023music}, which can enable conditional one-to-many-effect modeling \cite{chen2024towards, wright2024open} or serve as optimization objectives for inference-time effect matching \cite{steinmetz2024stito}. 
Another strategy estimates the dry signal with an effect removal model, trained on a diverse effect population, before applying supervised effect estimation \cite{hinrichs2024blind}.
While these methods succeed in modeling unseen effects, their performance depends heavily on the quality and diversity of pre-training data.
Generalization requires large, well-curated effect populations, and many methods suffer from flexibility constraints, being limited to simple audio effect implementations with few controllable parameters \cite{steinmetz2024stito, steinmetz2022style, peladeau2024blind, take2024audio} or fixed architectures dictated by conditional modeling paradigms \cite{wright2024open, chen2024towards}.

This work explores unsupervised approaches that rely only on unpaired examples from input and output data distributions, avoiding reliance on predefined effect populations and their associated generalization challenges.
We focus on methods that are agnostic to the functional form of the effect model, allowing them to optimize arbitrary operator models, including black-box and gray-box models, as long as they are suitable for modeling the system.
This flexibility is particularly advantageous for emulating real-world effects with unknown or highly complex behaviors,
and the optimized operator can serve as a computa\-tionally-cheap and real-time capable audio effect.
Specifically, we investigate two probabilistic frameworks that fulfill these criteria: one based on generative diffusion models,
which is novel in this context, and another based on adversarial training, which has been proposed previously \cite{wright2023adversarial}.

Diffusion models \cite{song2020score, karras2022elucidating} have emerged as state-of-the-art generative modeling techniques across various domains,
including audio generation \cite{liu2023audioldm, evans2025stable} and restoration \cite{moliner2024:zeroshot, lemercier2025diffusion}.
However, their utility in this work does not lie in their generative capabilities but in their potential to serve as data-driven priors for unsupervised system identification, a relatively unexplored application.
Recent studies have applied diffusion models to blind inverse problems \cite{moliner2023solving, laroche2024fast},
where the unknown degradation operator is optimized jointly with the clean signal estimate,
given only distorted measurements and a diffusion model trained on examples from the reference signal distribution. 
Such an approach has been explored in historical music restoration \cite{moliner2024:zeroshot, moliner2024diffusion} and speech dereverberation \cite{moliner2024buddy, lemercier2024unsupervised},
where \emph{linear} degradation operators were optimized as a~byproduct of the restoration process.
Building on previous findings \cite{svento2025NLDistortionDiff}, which demonstrated that diffusion-based methods can estimate a memoryless nonlinearity,
we now extend this methodology to general classes of \emph{nonlinear} operators in audio.

Methods based on adversarial training aim to align the output of a learned effect model with the target distribution using a discriminator model,
which is trained with an adversarial objective to the effect model. 
Wright et al.~\cite{wright2023adversarial} first proposed an adversarial approach for unsupervised guitar amplifier modeling with unpaired data, employing discriminators designed in the spectrogram domains. 
Chen et al.~\cite{chen2024:gan_extension} later extended this framework by experimenting with alternative discriminator architectures. 
Recently, Park et al.~\cite{park2025solving} applied adversarial training to optimize a~larger family of audio effects and degradation operators, though their method conditions on unpaired effected measurements and requires pre-training with a population of known audio effects, as in supervised approaches.

While adversarial approaches have demonstrated success in unsupervised effect estimation, they are known to suffer from training instabilities.
In particular, discriminator collapse can occur when the discriminator's training dynamics become unbalanced relative to the operator, a risk exacerbated by limited or unbalanced data availability \cite{thanh2020GANcollapse, pml2Book}. 
In many practical cases, one cannot assume access to a sufficiently large and diverse set of target-domain data, increasing the likelihood of instability and poor optimization outcomes.
By contrast, the diffusion-based approaches do not rely on adversarial training and are potentially less susceptible to mode collapse or training instabilities. 
This suggests that diffusion models may provide a more reliable solution for unsupervised nonlinear system identification, particularly in data-scarce scenarios.

This paper is structured as follows.
Sec.~\ref{sec:problem} formalizes the problem from a probabilistic perspective.
Sec.~\ref{sec:methods} introduces the two approaches under comparison,
describing their key principles.
Sec.~\ref{sec:operators} describes the operator models used in our experiments, including black-box models parameterized with neural networks, as well as a gray-box model based on a \ac{wh} structure with a novel design.
Sec.~\ref{sec:experiments} evaluates both methods in guitar distortion modeling, analyzing their behavior when a limited amount of target-domain data is available. The range of operator models that each method can optimize is explored, and their suitability for blind system identification in guitar distortion effects is assessed.
Additionally, we compare a \ac{wh} model to black-box models in this setting.
Finally, Sec.~\ref{sec:conclusions} summarizes our findings.

\section{Problem Formulation}\label{sec:problem}

Let \(\mathcal{X} = \{\mathbf{x}^{(m)}\}_{m=1}^{M}\) denote a dataset of source audio signals, where each \(\mathbf{x} \in \mathbb{R}^L\) is assumed to be drawn from a prior distribution \( p_x \). Additionally, let \( \mathcal{Z}_0 = \{\mathbf{z}_0^{(n)}\}_{n=1}^{N} \) represent an independent dataset of signals also drawn from the same distribution \( p_x \). Importantly, the dataset \(\mathcal{Z}_0\) is unobserved; instead, we are provided with a dataset \(\mathcal{Y} = \{\mathbf{y}^{(n)}\}_{n=1}^{N}\) of effected measurements.
Each observed signal \(\mathbf{y} \in \mathbb{R}^L\) is assumed to be generated from a~corresponding clean source signal \(\mathbf{z}_0 \) through an unknown distortion process, described by a function
\begin{equation}
\mathbf{y} = f(\mathbf{z}_0),
\end{equation}
where \( f: \mathbb{R}^L \to \mathbb{R}^L \) represents the distortion operator. The function \( f \) is assumed to be deterministic, time-invariant, and otherwise unknown. Our goal is to estimate it.

Since \( \mathbf{z}_0 \) follows the prior distribution \( p_x \), the distribution of \( \mathbf{y} \) is induced through the transformation \( f \). Moreover, since \( f \) is deterministic, the conditional distribution of the distorted measurements \( \mathbf{y} \) given the source signal \( \mathbf{z}_0 \) is expressed as a Dirac delta:
\begin{equation}\label{eq:likelihood_dirac}
p(\mathbf{y} | \mathbf{z}_0) = \delta(\mathbf{y} - f(\mathbf{z}_0)).
\end{equation}
This means that \( p_y \) is implicitly defined as  
\begin{equation}\label{eq:py}
p_y(\mathbf{y}) = \int_{\mathbb{R}^L} \!\delta(\mathbf{y} - f(\mathbf{z}_0)) p_x(\mathbf{z}_0) \,\mathrm{d}\mathbf{z}_0.
\end{equation}
Since \( f \) may not be invertible, different values of \(\mathbf{z}_0\) can map to the same \(\mathbf{y}\), leading to potential density transformations that are difficult to express in closed form.

\section{Unsupervised operator estimation} \label{sec:methods}

To approximate the unknown function \( f \), we introduce a parametric model \( \hat{f}(\cdot\, ; \psi) \), where \( \psi \) represents the learnable parameters. This model can take various functional forms, ranging from black box approaches, such as neural networks including \acp{cnn}, \acp{rnn}, or state space models, to gray box approaches, such as \ac{wh} models, which incorporate stronger inductive biases that reflect prior knowledge about the distortion process.
Our objective is to optimize \(\psi\) using the unpaired datasets \(\mathcal{X}\) and \(\mathcal{Y}\), i.e., in the absence of explicit supervision through paired samples.%

In this section, we explore two distinct approaches for tackling the unpaired estimation task. 
First, we introduce a novel method based on an \ac{em} objective using diffusion models, inspired by recent advances in image inverse problems \cite{laroche2024fast} and speech dereverberation \cite{lemercier2024unsupervised}. 
Second, we examine the adversarial approach proposed recently 
\cite{wright2023adversarial}, which formulates the problem as a generative adversarial learning task. 
These approaches use different strategies for estimating \( \hat{f}(\cdot\,; \psi) \), and are summarized in Fig.~\ref{fig:diagrams}.

In the diffusion-based approach of Fig.~\ref{fig:diagrams}(a), the source dataset \(\mathcal{X}\) is used to train a score model \(s_\theta\). Diffusion posterior sampling is applied to estimate the unseen variables in the set \(\mathcal{Z}_0\), from which the distorted dataset \(\mathcal{Y}\) originates. The operator parameters $\psi$ are optimized through EM updates, while the estimates of the elements from \(\mathcal{Z}_0\) are refined. 

In the adversarial approach of Fig.~\ref{fig:diagrams}(b), the source dataset \(\mathcal{X}\) is processed by the operator \(\hat{f}(\cdot\,; \psi)\), producing the estimated output \(\mathcal{\tilde{Y}}\). A discriminator \(D_\phi\) is trained to distinguish \(\tilde{\mathcal{Y}}\) from the distorted dataset \(\mathcal{Y}\), allowing the training of  \(\hat{f}(\cdot\,; \psi)\) through adversarial learning \cite{wright2023adversarial}.

\input{figures/diagrams_diffadv}

\subsection{Proposed Diffusion-Based Approach }

The first approach builds on recent advancements in diffusion models for blind inverse problems in music \cite{moliner2024blind, moliner2024diffusion}, speech \cite{lemercier2024unsupervised, nortier2024unsupervised}, and image \cite{laroche2024fast} restoration. These methods jointly estimate the degradation operator alongside the restored signal at inference time.  

\begin{sloppypar}
Following \cite{lemercier2024unsupervised, nortier2024unsupervised, laroche2024fast}, we formulate the operator estimation problem as an \ac{em} objective over an observed dataset \( \mathcal{Y} = \{\mathbf{y}^{(n)}\}_{n=1}^{N} \):  
\end{sloppypar}
\begin{equation} \label{eq:em}
    \underset{\psi}{\mathrm{max}} \,
    \mathbb{E}_{\mathcal{Z}_0\sim p(\mathcal{Z}_0 | \mathcal{Y})} \log p(\mathcal{Y} | \mathcal{Z}_0 ; \psi) \,,
\end{equation}
where \( \mathcal{Z}_0 = \{\mathbf{z}_0^{(n)}\}_{n=1}^{N} \) represents the latent clean source signals corresponding to \( \mathcal{Y} \), inferred during optimization.  

Instead of the theoretical Dirac likelihood in Eq.~\eqref{eq:likelihood_dirac}, we introduce a convex surrogate:  
\begin{equation}\label{eq:likelihood_psi}
    p(\mathcal{Y} | \mathcal{Z}_0; \psi) \propto
    \exp \left( - \zeta\sum_{n=1}^N \mathcal{C}(\mathbf{y}^{(n)}, \hat{f}(\mathbf{z}_0^{(n)}; \psi)) \right),
\end{equation}
where \( \mathcal{C}(\cdot, \cdot) \) denotes a convex cost function that quantifies the discrepancy between the observed and estimated signals, aggregated across the dataset, and $\zeta$ is a scaling hyperparameter.

The key challenge in optimizing Eq.~\eqref{eq:em} is evaluating the expectation, which requires drawing samples from the posterior distribution \( p(\mathcal{Z}_0 | \mathcal{Y}) \).
Since the dataset \( \mathcal{Y} \) consists of \( N \) distorted signals \( \mathbf{y}^{(n)} \), the algorithm requires estimating the corresponding source input signals \( \mathbf{z}_0^{(n)} \) for each \( n \in \{1, \dots, N\} \).
For clarity, we omit the index \( i \) in the following discussion, but the following procedure must be repeated for each observation in the dataset.

The estimation of each \(\mathbf{z}_0\) is performed using approximate posterior sampling with a diffusion model trained exclusively on the clean dataset \(\mathcal{X}\). The optimization procedure alternates between two steps, also visualized in Fig.~\ref{fig:diagrams}(a). In the E-step, given fixed parameters \(\psi\) and the trained diffusion model, we estimate the source samples \(\mathbf{z}_0\). In the M-step, using the estimated \(\mathbf{z}_0\), we update the parameters \(\psi\) of the degradation model to maximize the likelihood of the observations. 

\subsubsection{Diffusion models for posterior sampling} \label{sec:diff_inverse}

A diffusion model provides a powerful framework for sampling from complex distributions, such as \( p_x \), and for sampling from approximate posterior distributions, which factorize as
\( p(\mathbf{z}_0 | \mathbf{y}; \psi) \propto  p(\mathbf{y} | \mathbf{z}_0; \psi) p_x(\mathbf{z}_0) \), useful for solving inverse problems \cite{moliner2023solving, chung2022diffusion}. Diffusion models approach the generation problem by breaking it into a sequence of denoising steps. 
 The process starts from a Gaussian prior \( \mathbf{z}_T \sim \mathcal{N}(\mathbf{0}, \sigma_T^2 \mathbf{I}) \) and gradually refines the sample until it follows the target data distribution \( \mathbf{z}_0 \sim p_x \). This transformation is governed by the probability flow ordinary differential equation (ODE):  
\begin{equation}\label{eq:ode}
    \text{d}\mathbf{z}_\tau = - \tau  \nabla_{\mathbf{z}_\tau}\! \log p(\mathbf{z}_\tau) \text{d}\tau, 
\end{equation}
\begin{sloppypar}
\noindent where time \( \tau \) evolves backward from \( T \) to \( 0 \).
Since the score function \( \nabla_{\mathbf{z}_\tau}\!\log p(\mathbf{z}_\tau) \) is generally intractable, it is approximated using a time-conditional deep neural network \( s_\theta(\mathbf{z}_\tau, \tau) \approx \nabla_{\mathbf{z}_\tau}\! \log p(\mathbf{z}_\tau) \), trained via denoising score matching \cite{vincent2011connection} on the available dataset of clean signals \( \mathcal{X} \). 
We adopt the same diffusion parameterization as in previous works \cite{svento2025NLDistortionDiff, lemercier2024unsupervised} and refer to \cite{svento2025NLDistortionDiff} for a more detailed introduction.
\end{sloppypar}

To sample from the posterior, we replace the score function in Eq.~\eqref{eq:ode} with the posterior score~\cite{song2020score}:
\begin{equation}
    \nabla_{\mathbf{z}_\tau}\! \log p(\mathbf{z}_\tau | \mathbf{y}; \psi) =
    \nabla_{\mathbf{z}_\tau} \!\log p(\mathbf{z}_\tau )
    + \nabla_{\mathbf{z}_\tau } \!\log p(\mathbf{y} | \mathbf{z}_\tau; \psi).
\end{equation}
Following Chung et al.~\cite{chung2022diffusion}, we approximate this as
\begin{equation}
    \nabla_{\mathbf{z}_\tau} \!\log p(\mathbf{z}_\tau | \mathbf{y}; \psi) \approx
    s_\theta(\mathbf{z}_\tau, \tau)
       + \zeta(\tau) \nabla_{\mathbf{z}_\tau}\! \log
 p(\mathbf{y}|\hat{\mathbf{z}}_0(\mathbf{z}_\tau, \tau) ; \psi),
\end{equation}
where the prior score is replaced by \( s_\theta(\mathbf{z}_\tau, \tau) \), and the likelihood score is estimated using the one-step denoised estimate $\hat{\mathbf{z}}_0(\mathbf{z}_\tau) = \mathbb{E}[\mathbf{z}_0 | \mathbf{z}_\tau]$, which can be obtained without extra cost via Tweedie’s formula:
\begin{equation}
    \hat{\mathbf{z}}_0(\mathbf{z}_\tau, \tau)= \tau^2 s_\theta(\mathbf{z}_\tau, \tau) +\mathbf{z}_\tau.
\end{equation}

\subsubsection{E-Step}

 By discretizing and solving the posterior-replaced ODE introduced in Section \ref{sec:diff_inverse}, from $\tau=T$ to $\tau=0$, we could obtain samples from the approximate posterior \( p(\mathbf{z}_0 | \mathbf{y}; \psi) \), from which we could evaluate the expectation in Eq.~\eqref{eq:em}.  
However, simulating this conditional reverse diffusion process is computationally expensive due to the need for \( T \) evaluations of \( s_\theta \) per \ac{em} iteration.
To reduce this cost, we integrate \ac{em} updates with the ODE discretization, following \cite{laroche2024fast, lemercier2024unsupervised}. 
Specifically, we discretize the time variable \(\tau\) into \(K\) steps,  
\[
\{\tau_1, \dots, \tau_{k-1}, \tau_{k}, \dots, \tau_K \}, \quad \text{with} \quad \tau_1 = T \quad \text{and} \quad \tau_K \approx 0.
\]

At each step $k$, we approximate the expectation in Eq.~\eqref{eq:em} using an intermediate latent variable \(\mathbf{z}_{\tau_k}\):
\begin{equation} \label{eq:em_approx}
    \mathbb{E}_{\mathbf{z}_0\sim p(\mathbf{z}_0 | \mathbf{y}; \psi)} \log p(\mathbf{y} | \mathbf{z}_0 ; \psi) 
    \approx
    \mathbb{E}_{\mathbf{z}_0\sim p(\mathbf{z}_0 | \mathbf{z}_{\tau_k})} \log p(\mathbf{y} | \mathbf{z}_0 ; \psi).
\end{equation}
This approximation is motivated by the fact that the reverse diffusion process gradually refines samples toward the posterior mode. Since \(\mathbf{z}_{\tau_k}\) already encodes significant information about \(\mathbf{z}_0\), it serves as a useful intermediate representation. Rather than explicitly conditioning on \(\mathbf{y}\) at every \ac{em} iteration, we rely on the structure of the conditional reverse diffusion process to implicitly incorporate the data constraint.

\begin{sloppypar}
Following \cite{chung2022diffusion}, we further approximate \( p(\mathbf{z}_0 |\mathbf{z}_\tau) \) as a Dirac delta located at the one-step denoised estimate: $p(\mathbf{z}_0 |\mathbf{z}_\tau) \approx \delta(\hat{\mathbf{z}}_0(\mathbf{z}_\tau, \tau))$.
This results in the following one-sample Monte Carlo estimate of the expectation from Eq.~\eqref{eq:em}:
\end{sloppypar}
\begin{equation} \label{eq:em_approx_1}
    \mathbb{E}_{\mathcal{Z}_0\sim p(\mathcal{Z}_0 | \mathcal{Y})} \log p(\mathcal{Y} | \mathcal{Z}_0 ; \psi) 
    \approx
    \log p(\mathcal{Y} | \hat{\mathcal{Z}}^k_0 ; \psi) 
    \,,
\end{equation}
where $\hat{\mathcal{Z}}^k_0 = \{ \hat{\mathbf{z}}_0(\mathbf{z}_{\tau_k}^{(n)},\tau)  \}_{n=1}^N$. The latent variables \(\mathbf{z}_{\tau_{k+1}}^{(n)}\) are then updated following the process explained in Sec. \ref{sec:diff_inverse}.

\subsubsection{M-Step}
In the M-step, we update the operator parameters \( \psi \) to maximize the expected log-likelihood. 
By integrating Eqs.~\eqref{eq:likelihood_psi} and \eqref{eq:em_approx_1} into Eq.~\eqref{eq:em}, we obtain the M-step objective:
\begin{equation}
    \psi \leftarrow 
     \underset{\psi}{\mathrm{arg \, min}} \sum_{n=1}^{N} \mathcal{C}
    \left( \mathbf{y}^{(n)}\!,\, \hat{f}(\hat{\mathbf{z}}_0(\mathbf{z}_{\tau_k}^{(n)},\tau); \psi) \right),
\end{equation}
In practice, this objective is optimized via stochastic gradient descent. Optimization is performed by sampling random batches of pairs \( \{ \mathbf{y}, \hat{\mathbf{z}}_0 \} \) and updating the parameters accordingly.

\subsection{Adversarial Approach}

The second approach, sketched in Fig.~\ref{fig:diagrams}(b), is inspired by \acp{gan}  \cite{goodfellow2014generative} and was proposed for guitar amplifier modeling by Wright et al.~\cite{wright2023adversarial}. 
It consists of the following optimization objective \cite{pml2Book}:  
\begin{equation}
    \psi= \underset{\psi}{\text{arg min}} \, \mathcal{D}(
     p_y, 
     \hat{p}_y^\psi ),
\end{equation}  
where \( \hat{p}_y^\psi  \) denotes the distribution induced by $p_x$ through $\hat{f}(\cdot\,;\psi)$, defined analogously to Eq.~\ref{eq:py} for \( p_y \), and \( \mathcal{D}(\cdot, \cdot) \) denotes a distributional distance or divergence.
\( \mathcal{D}(\cdot, \cdot) \) is designed such that it attains its minimum when \( p_y = \hat{p}_y^\psi \).  

The distributional distance \( \mathcal{D}(\cdot, \cdot) \) is parameterized by a discriminator or critic \( D(\cdot\, ;\phi): \mathbb{R}^L \rightarrow \mathbb{R} \), typically a neural network with parameters \( \phi \), such that  
\begin{align}
\mathcal{D}(p_y, \hat{p}_y^\psi ) &= \underset{\phi}{\text{arg max}} 
\Bigg\{
\mathbb{E}_{\mathbf{y} \sim p_y} \big[ -\max(0, 1 - D(\mathbf{y};\phi)) \big]  \notag \\ 
&\quad + 
\mathbb{E}_{\mathbf{x} \sim p_x} \big[ -\max(0, 1 + D(\hat{f}(\mathbf{x}; \psi);\phi)) \big] 
\Bigg\}
\end{align}  
which corresponds to the hinge loss objective \cite{lim2017geometric}, as implemented in \cite{wright2023adversarial}. 
In practice, the expectations in the objective function are approximated using Monte Carlo estimates. 
The dataset of clean signals \( \mathcal{X} \) provides realizations of \( p_x \), while the dataset of distorted signals \( \mathcal{Y} \) serves as realizations of \( p_y \).  

\begin{sloppypar}
This formulation results in a minimax optimization game, where the discriminator \( D(\cdot \,;\phi) \) is trained adversarially against the operator \( \hat{f}(\cdot\,; \psi) \). The operator aims to transform \( p_x \) such that its output distribution aligns with \( p_y \), while the discriminator attempts to distinguish real samples from transformed ones.  One common issue in this adversarial framework is the discriminator collapse, where \( D(\cdot \,;\phi) \) becomes too strong and provides poor gradient information to $\hat{f}(\cdot\,; \psi)$, or conversely, too weak, failing to guide the operator’s learning process. 
\end{sloppypar}

\section{Operator models} \label{sec:operators}

The parametric model \( \hat{f}(\cdot\, ; \psi) \) can take various forms. In this paper, we explore different black-box models parameterized by neural networks, as well as a gray-box approach, specifically a \ac{wh} model.

\subsection{Black-Box Operator Models}

Black-box models are data-driven and do not require explicit know\-led\-ge of the underlying system in their design. Deep neural networks, as universal function approximators, are commonly used to approximate the input-output behavior of an effect.
In this study, we focus on two specific neural architectures that have proven successful at modeling audio effects: a \ac{gcn} and S4, a state-space model \cite{gu2022efficiently}.

The \ac{gcn} consists of a stack of temporally dilated convolutional operations combined with gated activation functions \cite{damskagg2019deep}. This architecture was used for unsupervised guitar effect estimation with an adversarial strategy \cite{wright2023adversarial}, and in this study, we use the same architecture as in \cite{wright2023adversarial}, consisting of 31k parameters.

S4, an architecture based on state-space models \cite{gu2022efficiently}, has been applied to nonlinear audio effects like dynamic range compression \cite{yin2024modelinganalogdynamicrange} and virtual analog effects \cite{simionato2024comparative}, often outperforming other architectures \cite{comunità2025differentiableblackboxgrayboxmodeling}. We use its implementation from \(\nabla\)Fx \cite{comunità2025nablafxframeworkdifferentiableblackbox}\footnote{\href{https://github.com/mcomunita/nablafx}{https://github.com/mcomunita/nablafx}}, with a 19k-parameter non-conditional configuration \cite{comunità2025differentiableblackboxgrayboxmodeling}.

\subsection{Wiener-Hammerstein Model}

\input{figures/wh3}

One of the most widely used approaches for modeling nonlinear systems is the \acl{wh} model, which represents the system as a serial connection of a \ac{lti} block, followed by a static nonlinearity, and another \ac{lti} block \cite{eichas2018gray, nercessian2021lightweight, colonel2022reverse, comunità2025differentiableblackboxgrayboxmodeling}.
Unlike black-box models, gray-box approaches such as \ac{wh} structures offer better interpretability of the learned transfer function, allowing a detailed analysis of each block after the optimization process. In the context of unsupervised optimization, we hypothesize that a more targeted model with a constrained parameter space, such as a \ac{wh} model, can be particularly beneficial.
The model's inductive biases, derived from prior knowledge of the unknown operator, can potentially help guide the optimization, improving stability and robustness. However, this advantage may come at the cost of reduced expressivity, as the model structure may limit its ability to fit highly complex systems that cannot be predicted with the utilized parametric structure.

The structure employed in this work is shown in Fig.~\ref{fig:wiener-hammerstein}. The \ac{lti} blocks are designed as equalizers implemented in the frequency domain, similar to the approach in \cite{colonel2021reverse}. The magnitude responses are optimized at a reduced grid, spaced according to third-octave bands. These values are then linearly interpolated to cover the entire frequency range. 
This is implemented by computing the \ac{stft} of the signal and multiplying each column (i.e., each time frame) element-wise with the interpolated frequency response.
Additionally, we optimize the phase values of the frequency-domain filter across the entire frequency range, allowing the operator to adapt to unknown phase responses.
For the \ac{stft}, a Hann window with 2048 samples and 75\% overlap is used, along with zero-padding to double the window length, preventing temporal aliasing.
The inverse \ac{stft} is posteriorly applied to recover the filtered waveform.

The static nonlinearity is parameterized with a \ac{ccr} spline, as proposed in previous work \cite{svento2025NLDistortionDiff}, since it outperformed alternative parameterizations such as Multilayer Perceptrons \cite{kuznetsov2020diffIIR} or parametric tanh structures \cite{colonel2022reverse} for modeling memoryless nonlinear distortion using a similar diffusion-based framework. The spline is parameterized with 41 control points.

\section{Guitar Distortion Modeling Experiments} \label{sec:experiments}

We evaluate the performance of the diffusion-based and adversarial methods in unsupervised guitar distortion modeling.
To do so, we use a similar experimental framework as \cite{wright2023adversarial}.
We investigate the effect of reducing the amount of available effected data in the results, reducing the size of $\mathcal{Y}$,  and how the two methods perform when different operator models are used. Audio examples are available on the webpage\footnote{\href{https://michalsvento.github.io/UnNAFx/}{https://michalsvento.github.io/UnNAFx/}}.

\begin{figure*}[t]
    \centering
    \includegraphics[ width=0.96\linewidth,trim={0 0 0 0.12cm}, clip]{figures/res_cos_dist_v2.pdf}
    \vspace{-0.45cm}
    \caption{Objective metrics AFx-Rep, $\ell_1\log\mathrm{MSS}$, and $\ell_1\mathrm{MSS}$ calculated on the test set, based on the amount of observed effected data. }
    \label{fig:res}
\end{figure*}

\subsection{Data}\label{exps.data}

Following \cite{wright2023adversarial}, we utilize the fourth subset of the IDMT-SMT Guitar dataset \cite{kehling2014automatic}, which comprises 64 short musical excerpts spanning various styles and tempos. While the dataset includes recordings from two different guitars, we exclusively use those from the Career SG guitar. Each recording is processed with three distinct distortion effects of increasing severity: ‘Clean Distortion,’ ‘Light Distortion,’ and ‘Heavy Distortion.’
The distortion effects were applied using commercial audio plugins, consistent with the procedure in \cite{wright2023adversarial}.
All the material is sampled at 44.1\,kHz.

\begin{sloppypar}
Our experiments require the following distinct, non-overlapping dataset splits: two unpaired datasets representing the input and output distributions, denoted as $\mathcal{X}$ and $\mathcal{Y}$, respectively, and a paired test set ($\mathcal{X}_\text{test}$, $\mathcal{Y}_\text{test}$) for evaluation. Of the 37 min of available recordings, we allocate 
16 min for $\mathcal{X}$, and 16 min for $\mathcal{Y}$. Additionally, we construct three reduced versions of $\mathcal{Y}$ containing 18 s, 1 min, and 4 min of distorted recordings.
The remaining 5~min are used for the test set, which is segmented into chunks of 6~s, resulting in 50 segments.
Our dataset split differs from the one used in \cite{wright2023adversarial}, as this configuration enables a more systematic evaluation of both approaches. Consequently, our results may not be directly comparable to those reported in \cite{wright2023adversarial}.
\end{sloppypar}

\subsection{Experimental Details: Diffusion-Based Approach}  \label{exp.diffusion}

Our diffusion-based method closely follows the configuration of previous work \cite{svento2025NLDistortionDiff}, using the same hyperparameter choices unless otherwise specified.
The diffusion model operates in the waveform domain, while its score network is built on an architecture based on the Constant-Q Transform (CQT), as proposed in \cite{moliner2024diffusion}. This design leverages the invertibility and differentiability of the CQT to introduce inductive biases tailored to musical signals, while retaining the flexibility of waveform-domain modeling.
The model is trained on 6-second audio segments.
The model was pre-trained for 80k iterations using the EGDB dataset \cite{chen2022towards}, followed by 14k iterations of training on the dataset split $\mathcal{X}$, with a batch size of 4. The training phase (excluding pre-training) took 70 min\footnote{All computation times here were measured on an NVIDIA H200 GPU.}.
While the impact of pre-training remains to be formally evaluated, preliminary experiments suggest that it may have only a minimal effect.

We use a consistent hyperparameter configuration for all the operator optimization experiments, setting $T=101$ steps, which corresponds to both the reverse diffusion discretization and the number of EM iterations. 
The likelihood scaling parameter $\zeta$ in Eq.~\eqref{eq:likelihood_psi} is defined following \cite{svento2025NLDistortionDiff} as a function of $\tau$ controlled by $\tilde{\zeta}=0.2$. 
 As the cost function $C(\cdot, \cdot)$, we employ a $\ell_2$-norm in a compressed STFT representation \cite{moliner2024buddy,lemercier2024unsupervised}, using a compression factor of 0.5.
This compression equalizes the spectral energy distribution and enhances high-frequency content,
which typically has lower energy but is perceptually important.
Each M-step includes 20 operator optimization steps. We employ the AdamW optimizer with random batches of 4 examples per iteration, a learning rate of 0.001,
momentum parameters ($\beta_1=0.9$, $\beta_2=0.99$), and a weight decay of 0.01.

Operator optimization times varied based on dataset $\mathcal{Y}$ size: approximately 1 h for 16 min of data, 18 min for 4 min of data, 6~min for 1 min, and 3 min for 18 s. Exact times may vary depending on operator efficiency and implementation-specific details such as logging.
Certain operations during the E-step and latent variable updates, particularly the forward and backward score model evaluations, demand substantial memory if processed naively in a single evaluation. To mitigate memory overhead, we process all dataset elements in small batches of 1, 2, or 4, depending on available GPU memory.

\subsection{Experimental Details: Adversarial Approach} \label{exp.adversarial}

Following \cite{wright2023adversarial}, we adopt the set of three log-mel spectrogram discriminators, each operating on 160 mel bands but with varying window sizes of 512, 1024, and 2048 samples. This configuration was chosen as it demonstrated superior performance in the experiments reported in that study.
All models were trained for exactly 100k iterations with batches of 5 segments of 1.5\,s each, requiring approximately 2 h.
We follow the remaining hyperparameter settings from \cite{wright2023adversarial}.
Experiments using the \ac{wh} model proved unstable with this approach, suffering from severe discriminator collapse. Consequently, they were excluded from the evaluation, and only the black-box models, GCN and S4, were studied.

We acknowledge the extensive body of work aimed at improving the stability of adversarial training \cite{miyato2018spectral, gulrajani2017improved}, and recognize that some of these techniques could be relevant to our setting, particularly in imbalanced scenarios. However, exploring and properly implementing these approaches is considered beyond the scope of the present work. Therefore, we limit ourselves to the configuration proposed in \cite{wright2023adversarial}.

\subsection{Experimental Details: Supervised Baseline}

In addition to the unsupervised methods that are the focus of this study, we incorporate a supervised baseline to serve as an upper performance bound for the unsupervised approaches. This baseline is trained using the dataset $\mathcal{Y}$, with reduced size when applicable, along with the paired input signal, which is unavailable in the unsupervised setting. The supervised operator models were trained for 5k iterations using the Adam optimizer with a learning rate of 0.0001 
and 6-s-long audio segments.

\subsection{Evaluation}

To systematically evaluate the performance of the different methods, we apply each optimized operator to all instances in the test set \(\mathbf{x}_\text{test} \in \mathcal{X}_\text{test}\), obtaining the corresponding signal estimates: \( \hat{\mathbf{y}} = \hat{f}(\mathbf{x}_\text{test}; \psi)\).
We then assess the quality of these estimates by comparing them to the paired ground-truth targets \(\mathbf{y} \in \mathcal{Y}_\text{test}\) using three objective evaluation metrics.

The first metric is based on the AFx-Rep representation \cite{ steinmetz2024stito}, which was specifically designed to capture information related to audio production style. 
To quantify the similarity between the reference signal $\mathbf{y}$ and its estimate $\hat{\mathbf{y}}$, we compute the cosine distance between their respective embeddings:
\begin{equation}
\text{dist}(\hat{\mathbf{y}}, \mathbf{y}) = 1 -
\frac{g(\hat{\mathbf{y}}) \cdot g(\mathbf{y})}
{\max(\|g(\hat{\mathbf{y}})\| \|g(\mathbf{y})\|, \epsilon)},
\end{equation}
The operator \( g \) is the AFx-Rep trained encoder,  $\cdot$ denotes the dot product, the norm used is the $\ell_2$ norm, and $\epsilon$ is a small constant for numerical stability.

Next, we compute the \(\ell_1\) distance between Multi-Scale Spectrograms (MSS), which measures the difference between \ac{stft} magnitudes across multiple analysis settings.
The \ac{stft} is computed using predefined window lengths $W$ = \{2048, 1024, 512, 256, 128, 64\}.
For each \( w \in W \), the number of FFT bins is set to \( w \), and the hop size is chosen as \( w / 4 \). 
The \(\ell_1\) MSS metric
is then defined as:  
\begin{equation}
  \ell_1 \textrm{MSS} (\hat{\mathbf{y}}, \mathbf{y}) =  \sum_{w \in W} \|\, |\mathrm{STFT}_w(\hat{\mathbf{y}})| - |\mathrm{STFT}_w(\mathbf{y})|\, \|_1,
\end{equation}
where $\mathrm{STFT}_w(\cdot)$ is the \ac{stft} operator with an analysis window of size $w$.
To obtain values on a more interpretable scale, we normalize the \(\ell_1\)-norm by taking its mean.

As the third metric, we define a logarithmic variant of the previously mentioned metric, \(\ell_1\! \log \mathrm{MSS}\), where \(\log_{10}\) is applied to the STFT magnitudes before computing the distance.

\subsection{Results}

The results of the objective metrics are shown in Fig.~\ref{fig:res}. As expected, the supervised experiments (green markers) consistently achieve the best performance across all scenarios, establishing a lower bound on the error that can be expected from the unsupervised methods.

When evaluating the robustness of unsupervised methods in data-scarce scenarios, it can be  observed that the adversarial me\-thod (blue markers in Fig.~\ref{fig:res}) exhibits a notable drop in performance as the duration of available data decreases.
In contrast, the diffusion-based approach (red markers in Fig.~\ref{fig:res}) maintains a more stable performance, even with as little as 18\,s of data, although a~slight decline is still noticeable.

\begin{sloppypar}
The experiments with the diffusion-based method outperform the adversarial ones in the `Clean' and `Light' distortion scenarios when only 18~s or 1 min of effected recordings are available, and achieves a comparable performance when 4 or 16 min are used.
In terms of AFx-Rep dist. (Fig.~\ref{fig:res}a,b) and 
\(\ell_1\! \log \mathrm{MSS}\)
(Fig.~\ref{fig:res}d,e), the adversarial approach yields slightly better results when the full dataset is available.
Conversely, the diffusion-based method achieves consistently lower errors on the linear
\(\ell_1\!\,\mathrm{MSS}\)
metric (Fig.~\ref{fig:res}g,h).
These differences are potentially attributed to the alignment between each method's training objective and the evaluation metrics. The adversarial model was trained using log-scaled mel-spectrogram features, which are more closely related to the
\(\ell_1\! \log \mathrm{MSS}\) 
metric. In contrast, the diffusion-based model was optimized using magnitude-compressed STFT features, which may be more aligned with linear-scale $\ell_1\,\mathrm{MSS}$, though the correspondence is not exact.
\end{sloppypar}

These trends differ in the `Heavy' distortion setting, where the adversarial method experiments consistently obtained better results than diffusion-based ones in terms of AFx-Rep dist (Fig.~\ref{fig:res}(c)) and $\ell_1$ log MSS (Fig.~\ref{fig:res}(f)).
Interestingly, this is not the case in terms of $\ell_1$ MSS, where the diffusion-based approach still obtains lower values.
Also in this case, for both unsupervised approaches, increased data availability almost consistently leads to improved operator estimation.

In the Heavy distortion setting, the trend shifts: the adversarial method consistently outperforms the diffusion-based approach in AFx-Rep distance (Fig.~\ref{fig:res}c) and
$\ell_1 \log \mathrm{MSS}$
(Fig.~\ref{fig:res}f). The diffusion-based method still performs better in terms of linear $\ell_1 \, \textit{MSS}$.  Also in this case, for both unsupervised approaches, performance generally improves with increased data availability.

\begin{sloppypar}
We do not observe a clear trend regarding the most suitable black-box model architecture: both GCN and S4 yield comparable performance when used with either of the unsupervised methods. While one may outperform the other in specific cases, the difference is not consistent across settings.  
Within the diffusion-based framework, a comparison between black-box and grey-box operators shows that the \ac{wh} model performs on par with both the GCN and S4 models (see Fig.~\ref{fig:res}), and slightly better in some cases. Attempts to train the \ac{wh} model using the adversarial approach were unsuccessful, and the corresponding results are therefore not included in the figure.  
These findings suggest that diffusion-based methods offer stable performance across different operator architectures, whereas the adversarial approach appears more sensitive to the choice of model.
\end{sloppypar}

\section{Conclusions}\label{sec:conclusions}

\begin{sloppypar}
This study addressed unsupervised operator estimation, with experiments on guitar distortion modeling, comparing diffusion-based and adversarial approaches. While adversarial methods performed well under heavy distortion, they lack consistency across scenarios. In contrast, diffusion-based methods show strong robustness to data scarcity and operator choice, making them a more reliable option overall. Additionally, diffusion methods offer the benefit of jointly estimating the clean guitar signal, although reconstruction quality has not been explored in this work.

Both approaches—diffusion-based and adversarial—require substantial training time and computational resources, which can limit their accessibility for applications that require training with low-power or time constraints.
A key limitation of the diffusion-based method is the need for a separately pre-trained model on clean guitar signals, and obtaining such dry data can be challenging depending on the context.
Future work should investigate how performance varies with different amounts of available dry data and assess the trade-offs between robustness and computational efficiency. 
Although all evaluated models are technically capable of real-time operation~\cite{wright2020real}, the practical computational demands remain an open area for further study.
Finally, although this study focused specifically on distortion effects, we believe the proposed framework could extend to other nonlinear audio effects, such as dynamic range compression or modulation effects, given the generality of the black-box operators used in training.

\end{sloppypar}

\section{Acknowledgments}
Part of this work was conducted during E.~Moliner's research visit to Brno University of Technology in March 2025. M.~Švento completed a four-month Erasmus+ traineeship at the Aalto Acoustics Lab in fall 2024. The authors acknowledge the computational resources provided by the Aalto Science-IT project. The work carried out at the Brno University of Technology was supported by the Czech Science Foundation (GA\v{C}R) Project No.\,23-07294S.

\input{DAFx25_arxiv_v2.bbl} %

\begin{onecolumn}

\appendix

\begin{algorithm*}[t]
\caption{Diffusion-based approach for operator estimation
}
\label{alg:diffusion}

\begin{algorithmic}
\Require Source dataset $\mathcal{X}=\{
\mathbf{x}^{(m)}
 \}_{m=0}^M$, effected dataset $\mathcal{Y}=\{
\mathbf{y}^{(n)}
 \}_{n=0}^N$

\State Train score model $\mathbf{s}_\theta$ using $\mathcal{X}$ via denoising score-matching  \Comment{Train diffusion model on clean data}

\State Initialize operator parameters $\psi$ %
\State {Sample $\mathcal{Z}_{\tau_1}\sim \{ \mathcal{N}(\mathbf{0},\sigma^2(\tau_1)\mathbf{I})
\}_{n=1}^N
$}
\Comment{Initialize latent variables with Gaussian noise}
\For{$k \leftarrow 1, \dots, K$}  \Comment{Expectation-Maximization (EM) iterations} 
    \State \textit{Phase~1~--~E-Step}%
    
    \State  $
    \hat{\mathcal{Z}}_0^k
    \leftarrow 
    \{
    \tau_k^2 s_\theta(\mathbf{z}_{\tau_k}^{(n)}, \tau_k) +\mathbf{z}_{\tau_k}^{(n)}
    \}_{n=0}^N
    $ \Comment{Compute one-step denoised estimates via Tweedie's formula}
    
    \State $\mathcal{L}(\psi; \mathcal{Y}, \hat{\mathcal{Z}}^k_0)  := -\sum_{n=1}^{N} \mathcal{C}
    \left( \mathbf{y}^{(n)}, \hat{f}(\hat{\mathbf{z}}_0(\mathbf{z}_{\tau_k}^{(n)},\tau); \psi) \right),
    $ \Comment{Log-likelihood approximation functional} 

    \State
    \State \textit{Phase~2~--~M-Step}%
    
    \For{$m \leftarrow 0, \dots, M_\text{its.}$}  \Comment{Optimize operator parameters}
        \State Select a random batch $(\mathcal{Y}_b, \hat{\mathcal{Z}}^k_{0,b}) \subset (\mathcal{Y}, \hat{\mathcal{Z}}^k_{0})$   \Comment{Sample paired mini-batch}
        \State $\psi \leftarrow \psi - \mathrm{Adam}
        \left( \nabla_\psi -\mathcal{L}(\psi, \mathcal{Y}_b, \hat{\mathcal{Z}}^k_{0,b})
        \right)$ 
        \Comment{Gradient update with Adam optimizer}
    \EndFor
    \State
    \State \textit{Phase~3~--~Latent variable update}%

    \For{$i \leftarrow 1, \dots, N$}  \Comment{Iterate over dataset; can be parallelized if sufficient memory is available}
    \State $\mathrm{d}\tau \leftarrow (\tau_{k+1}-\tau_k)$
    \State $\mathbf{z}^{(n)}_{\tau_{k+1}} \leftarrow {\mathbf{z}}^{(n)}_{\tau_k} - {\tau}_k \left(\mathbf{s}_\theta(\mathbf{z}^{(n)}_{\tau_k}, \tau_k) + \zeta(\tau)\nabla_{\mathbf{z}^{(n)}_{\tau_k}} \mathcal{L}
    \left(
    \psi; \mathbf{y}^{(n)}, \hat{\mathbf{z}}_0(\mathbf{z}^{(n)}_{\tau_k},\tau)
    \right) \right) \mathrm{d}\tau$ 
    \Comment{Update latent variables using DPS. The score function and the denoised estimates computed in the E-step  can be reused here to reduce computational cost.}
    \EndFor
\EndFor
\State \Return  $\hat{f}(\cdot, \psi)$ \Comment{Return the estimated operator}
\end{algorithmic}
\end{algorithm*}

\begin{algorithm*}[t]
\caption{Adversarial approach for operator estimation
}
\label{alg:adversarial}

\begin{algorithmic}
\Require Source dataset $\mathcal{X}=\{
\mathbf{x}^{(m)}
 \}_{m=0}^M$, effected dataset $\mathcal{Y}=\{
\mathbf{y}^{(n)}
 \}_{n=0}^N$

\State Initialize operator parameters $\psi$ %
\For{$k \leftarrow 1, \dots, K$}  \Comment{Adversarial training iterations} 
    \State \textit{Phase~1~--~Discriminator Update}%

    \State Sample unpaired random batches  $\mathcal{X}_b \subset \mathcal{X}$ , $\mathcal{Y}_{b^\prime} \subset \mathcal{Y}$
    \State $\mathcal{\tilde{Y}} \leftarrow 
    \{
    \hat{f}(\mathbf{x}^{(m)}; \psi)
    \}_{m=0}^B
    $
    
    \State $\phi
    \leftarrow
    \phi -
    \mathrm{Adam}
    \left(
    \nabla_\phi
    \left[
    \sum_{n=0}^{B^\prime} 
 \max(0, 1 - D(\mathbf{y}^{(n)}; \phi))
+ 
    \sum_{m=0}^{B} 
 \max(0, 1 + D(\tilde{\mathbf{y}}^{(m)}; \phi)) 
 \right]
\right)
    $ \Comment{Optimize Discriminator}

    \State
    \State \textit{Phase~2~--~Operator Update}%
    
    \State Sample new random batch  $\mathcal{X}_{b^{\prime \prime}} \subset \mathcal{X}$
    \State $\psi \leftarrow \psi - \mathrm{Adam}
    \left(
    \nabla_\psi
    \left[
    \sum_{m=0}^{B^{\prime \prime}}
    -D(
    \hat{f}(
   \mathbf{x}^{(m)};\psi
    );\phi
    )
    \right]
    \right)
    $ \Comment{Optimize operator}

\EndFor
\State \Return  $\hat{f}(\cdot, \psi)$ \Comment{Return the estimated operator}
\end{algorithmic}
\end{algorithm*}

\end{onecolumn}

\end{document}

%% file: figures/diagrams_diffadv.tex
\definecolor{cb1}{HTML}{D81B60}
\definecolor{cb2}{HTML}{1E88E5}
\definecolor{cb3}{HTML}{D29E02}
\definecolor{cb4}{HTML}{004D40}
\definecolor{cb5}{HTML}{864dbf}

\tikzstyle{myellipse} = [ellipse, draw, dotted, fill=white, inner sep=0pt, minimum height=25pt, minimum width=40pt]

\tikzstyle{mycircle} = [circle, draw, fill=white, inner sep=0pt, minimum size=20pt]
\tikzstyle{mycircle_dotted} = [circle, draw, dotted, fill=white, inner sep=0pt, minimum size=20pt]
\tikzstyle{mysquare} = [rectangle, draw, fill=black, inner sep=0.1pt, minimum width=35pt, minimum height=35pt, align=center]
\tikzstyle{mybranch} = [circle, draw, fill=black, inner sep=0pt, , minimum size=2pt]
\tikzstyle{myrectangle1} = [rectangle, draw, fill=cb1!20, inner sep=3pt, minimum width=30pt, minimum height=20pt, align=center]
\tikzstyle{myrectangle2} = [rectangle, draw, fill=cb2!20, inner sep=3pt, minimum width=30pt, minimum height=20pt, align=center]
\tikzstyle{myrectangle4} = [rectangle, draw, fill=cb3!20, inner sep=3pt, minimum width=30pt, minimum height=20pt, align=center]
\tikzstyle{myrectangle3} = [rectangle, draw, fill=cb4!20, inner sep=3pt, minimum width=30pt, minimum height=20pt, align=center]
\tikzstyle{myrectangle5} = [rectangle, draw, fill=cb5!20, inner sep=3pt, minimum width=30pt, minimum height=20pt, align=center]

\tikzstyle{myrectangle_white} = [rectangle, draw, fill=white, inner sep=3pt, minimum width=10em, minimum height=20pt, align=center]

\tikzstyle{sum} = [
  circle,
  draw,
  minimum size=9pt,
  append after command={
    \pgfextra{
      \draw (\tikzlastnode.north) -- (\tikzlastnode.south);
      \draw (\tikzlastnode.west) -- (\tikzlastnode.east);
    }
  },
]

\tikzstyle{product} = [
  circle,
  draw,
  minimum size=12pt,
  append after command={
    \pgfextra{
      \draw (\tikzlastnode.north east) -- (\tikzlastnode.south west);
      \draw (\tikzlastnode.north west) -- (\tikzlastnode.south east);
    }
  },
]

\tikzstyle{convolutioncircle} = [
  circle,
  draw,
  minimum size=24pt,
]

\newcommand{\specwidth}{1.35cm}
\newcommand{\sepnodex}{0.5cm}
\newcommand{\sepnodey}{0.4cm}
\newcommand{\sepnodexspec}{0cm}
\newcommand{\sepnodeyspec}{0.25cm}
\newcommand{\sepstate}{0.75cm}
\definecolor{figblue}{HTML}{154c79}
\definecolor{figgreen}{HTML}{a8db33}
\definecolor{figgrey}{HTML}{797979}

\begin{figure}[t]

\centering

\begin{subfigure}[b]{\linewidth}

\vspace{-0.1cm}
\hspace{1cm}
\begin{tikzpicture}

\coordinate (stateX) at (0,0.0);
\coordinate (stateY) at (4,-1.5);
\coordinate (stateZ) at (0, -1.5);
\coordinate (stateYhat) at (4.0, 0.0);
\coordinate (loss) at (2.5, -4.0);

\coordinate (pos_disc) at (6.0, -0.5);

\node[mycircle] (sx) at (stateX) {$\mathcal{X}$};

\node[mycircle_dotted] (sz) at (stateZ) {$\mathcal{Z}_0$};
\node[mycircle] (sy) at (stateY) {$\mathcal{Y}$};

\node[myrectangle3, align=center, right=1cm of sz] (f) {$\hat{f}(\cdot\,; \psi)$};

\node[myrectangle1, align=center, right=1cm of sx] (score) {$s_\theta$};

\draw[->, -Stealth] (sz) to (f);
\draw[->, -Stealth] (f) to (sy);

\draw[->, -Stealth, cb1!100, dotted, line width=0.3mm, out=0, in=180] (sx.east) to (score.west);

\draw[->, -Stealth, cb3!100, dotted, line width=0.3mm, out=200, in=90]     (score.west) ++(0, -0.15)  to (sz.north) ;

\draw[->, -Stealth, cb3!100, dotted, line width=0.3mm, out=170, in=60] (sy.north) ++(-0.3, -0.1) to (0.12, -1.10);

\draw[->, -Stealth, cb3!100, dotted, line width=0.3mm, out=120, in=40]     (f.north) to (0.25, -1.25) ;

\draw[->, -Stealth, cb4!100, dotted, line width=0.3mm, out=320, in=220] (sz.south) to (f.south);

\draw[->, -Stealth, cb4!100, dotted, line width=0.3mm, out=220, in=320] (sy.south) to (f.south);

\end{tikzpicture}
 \caption{\normalsize \textit{Proposed diffusion-based approach}}

\end{subfigure}

 \vspace{0.5cm}
\begin{subfigure}[b]{\linewidth}

\hspace{1cm}
\begin{tikzpicture}
\coordinate (stateX) at (0,0.0);
\coordinate (stateY) at (4,-1.5);
\coordinate (stateZ) at (0, -1.5);
\coordinate (stateYhat) at (4.0, 0.0);
\coordinate (loss) at (2.5, -4.0);

\coordinate (pos_disc) at (5.5, -0.75);

\node[mycircle] (sx) at (stateX) {$\mathcal{X}$};
\node[mycircle_dotted] (syhat) at (stateYhat) {$\mathcal{\tilde{Y}}$};

\node[mycircle] (sy) at (stateY) {$\mathcal{Y}$};

\node[myrectangle3, align=center, right=1cm of sx] (f) {$\hat{f}(\cdot\,; \psi)$};

\draw[->, -Stealth] (sx) to (f);
\draw[->, -Stealth] (f) to (syhat);

\node[myrectangle2, align=center] (Disc) at (pos_disc) {$D(\cdot;\phi)$};

\draw[->, -Stealth, cb4!100, dotted, line width=0.3mm, out=180, in=280] (Disc.west) to (f.south);

\draw[->, -Stealth, cb2!100, dotted, line width=0.3mm, out=0, in=250] (sy.east) to (Disc.south);

\draw[->, -Stealth, cb2!100, dotted, line width=0.3mm, out=0, in=110] (syhat.east) to (Disc.north);

\end{tikzpicture}
 \caption{\normalsize \textit{Adversarial approach \cite{wright2023adversarial}}}

\end{subfigure}

\vspace{-0.2cm}
    \captionsetup{font=small}
\caption{\textit{
High-level diagrams of the two studied unpaired operator estimation methods, (a)~the diffusion-based approach and (b)~the adversarial approach. 
Solid lines indicate the signal processing flow, while dashed lines represent optimization dependencies between components.
}}

\label{fig:diagrams}

\label{fig:training}

\end{figure}

%% file: figures/wh3.tex
\tikzstyle{product} = [
  circle,
  draw,
  minimum size=12pt,
  append after command={
    \pgfextra{
      \draw (\tikzlastnode.north east) -- (\tikzlastnode.south west);
      \draw (\tikzlastnode.north west) -- (\tikzlastnode.south east);
    }
  },
]

\tikzstyle{myrectangle1} = [rectangle, draw, fill=cb1!20, inner sep=3pt, minimum width=30pt, minimum height=20pt, align=center]
\tikzstyle{myrectangle2} = [rectangle, draw, fill=cb2!20, inner sep=3pt, minimum width=30pt, minimum height=20pt, align=center]
\tikzstyle{myrectangle4} = [rectangle, draw, fill=cb3!20, inner sep=3pt, minimum width=30pt, minimum height=20pt, align=center]
\tikzstyle{myrectangle3} = [rectangle, draw, fill=cb4!20, inner sep=3pt, minimum width=30pt, minimum height=20pt, align=center]
\tikzstyle{myrectangle5} = [rectangle, draw, fill=cb5!20, inner sep=3pt, minimum width=30pt, minimum height=20pt, align=center]

\begin{figure}[t!]
    \centering
    \begin{tikzpicture}[
        block/.style={draw, minimum width=1.5cm, minimum height=1cm, align=center, font=\footnotesize},
        node distance=0.8cm
    ]

        \node (x) { \( \mathbf{x}\)};

        \node[myrectangle4, right=0.5cm of x, inner sep=2pt] (geq1) {
             STFT-EQ \\ \vspace{-0.1cm}  \\
             
            \begin{tikzpicture}[scale=0.2]
                \draw[thick] (0,0) -- (1,0.8) -- (2,1.4) -- (3,0.5) -- (4,1.2) -- (5,0.9) -- (6,1.5) -- (7,0); %
            \end{tikzpicture}
        };

        \node[above=0.31cm of geq1] (lti1) {\textit{LTI}};
        
        \node[myrectangle3, right=0.5cm of geq1] (spline) {
             CCR Spline\\ 
            \begin{tikzpicture}[scale=0.25]
                \draw[thick, smooth] (-2,-1.5) .. controls (-1.5,-1.5) and (-1,-1) .. (0,0) 
                                     .. controls (1,1) and (1.5,1.5) .. (2,1.8)
                                     .. controls (2.5,1.9) and (3,2) .. (3.5,2);
                    
                \foreach \x/\y in {-2/-1.5, -1/-1, 0.5/0.5, 2/1.8, 3.5/2}
                    \draw (\x,\y) circle (4pt);
            \end{tikzpicture}
        };
        \node[above=0.01cm of spline] (nonlin) {\textit{Static nonlinearity}};
        
        \node[myrectangle4, right=0.5cm of spline, inner sep=2pt] (geq2) {
            STFT-EQ \\ \vspace{-0.1cm}  \\
            \begin{tikzpicture}[scale=0.2]
                \draw[thick] (0,0) -- (1,1) -- (2,0.6) -- (3,1.3) -- (4,0.7) -- (5,1.4) -- (6,0.9) -- (7,0); %
            \end{tikzpicture}
        };
        \node[above=0.31cm of geq2] (lti2) {\textit{LTI}};

        \node[right=0.5cm of geq2] (y) { \( \mathbf{y} \)};

        \draw[->,-Stealth] (x) -- (geq1);
        \draw[->,-Stealth] (geq1) -- (spline);
        \draw[->,-Stealth] (spline) -- (geq2);
        \draw[->,-Stealth] (geq2) -- (y);
    \end{tikzpicture}
    \caption{\textit{Structure of the employed Wiener-Hammerstein model.  }}
    \label{fig:wiener-hammerstein}
\end{figure}